\documentclass[conference]{IEEEtran}
\PassOptionsToPackage{}{hyperref}
\PassOptionsToPackage{obeyspaces}{url}

\usepackage{float, enumerate, xcolor, algorithm, algorithmic, tabto, balance, amsmath, graphicx, epstopdf, hyperref, multirow, dblfloatfix, textcomp, listings, eurosym, lstautogobble, multirow, tikz}
\usepackage{amsfonts}
\usepackage{comment}
\usepackage{amssymb}
\usepackage{booktabs}
\usepackage{url}
\usepackage{amssymb}
\usepackage[numbers,sort&compress]{natbib}
\usepackage[flushleft]{threeparttable}
\usepackage{subfigure}
\usetikzlibrary{arrows,backgrounds,positioning}
\usepackage{chngcntr}
\usepackage[overload]{empheq}
\usepackage{soul}
\usepackage[multiple]{footmisc} 

\newcommand{\ankit}[1]{\textcolor{black}{#1}}

\usepackage{array}
\newcolumntype{L}{>{\centering\arraybackslash}m{.5\columnwidth}|}

\usepackage{url}

\usepackage{pifont}% http://ctan.org/pkg/pifont
\usetikzlibrary{arrows,automata}

\graphicspath{ {./images/} }

\begin{document}
	\title{On the Feasibility of Profiling Electric Vehicles through Charging Data\vspace{-.35em}}
	
	\author{
		\IEEEauthorblockN{
			Ankit Gangwal\IEEEauthorrefmark{2}, 
			Aakash Jain\IEEEauthorrefmark{2}, and 
			Mauro Conti\IEEEauthorrefmark{4}
		}
	\thanks{This is an extended version of our paper in VehicleSec 2023~(co-located with NDSS Symposium 2023), San Diego, CA.}
		\IEEEauthorblockA{
			\IEEEauthorrefmark{2} International Institute of Information Technology, Hyderabad\\
			\IEEEauthorrefmark{4} University of Padua\\
		Email: gangwal@iiit.ac.in, aakash.jain@students.iiit.ac.in, mauro.conti@unipd.it}
	}
	
	\IEEEoverridecommandlockouts
	\makeatletter\def\@IEEEpubidpullup{6.5\baselineskip}\makeatother
	\IEEEpubid{\parbox{\columnwidth}{
			Symposium on Vehicles Security and Privacy (VehicleSec) 2023 \\
			27 February 2023, San Diego, CA, USA \\
			ISBN 1-891562-88-6 \\
			https://dx.doi.org/10.14722/vehiclesec.2023.23021 \\
			www.ndss-symposium.org
	}
	\hspace{\columnsep}\makebox[\columnwidth]{}}
	\maketitle

	\begin{abstract}
		Electric vehicles (EVs) represent the long-term green substitute for traditional fuel-based vehicles. To encourage EV adoption, the trust of the end-users must be assured. 
		\par
		In this work, we focus on a recently emerging privacy threat of profiling and identifying EVs via the analog electrical data exchanged during the EV charging process. The core focus of our work is to investigate the feasibility of such a threat at scale. To this end, we first propose an improved EV profiling approach that outperforms the state-of-the-art EV profiling techniques. Next, we exhaustively evaluate the performance of our improved approach to profile EVs in real-world settings. In our evaluations, we conduct a series of experiments including 25032 charging sessions from 530 real EVs, sub-sampled datasets with different data distributions, etc. Our results show that even with our improved approach, profiling and individually identifying the growing number of EVs \ankit{appear extremely difficult} in practice; at least with the analog charging data utilized throughout the literature. We believe that our findings from this work will further foster the trust of potential users in the EV ecosystem, and consequently, encourage EV adoption.
	\end{abstract}

\section{Introduction}
The growing concerns related to the climate crisis have led a global movement to adopt green and renewable energy for a sustainable future. Electric Vehicles~(EVs) represent a long-term ecological substitute for fossil fuel-based vehicles. EVs are even perceived as the key patrons for achieving near zero carbon footprint~\cite{evex1, evs1_20}. Today, EVs are becoming increasingly popular as well as gaining widespread adoption. As a representative example, the global sales of EVs in Q1~'21 were over 2.5 times of their sales in Q1 '20~\cite{evex40}. As an estimate~\cite{evs1_6}, the annual EV sales will reach over 31.1 million by 2030; which will represent approximately 32\% of new car sales worldwide. Furthermore, some vehicle manufactures have plans to produce only EVs by 2040~\cite{evex36}.
\par
With the increasing adoption of EVs, the demand for their charging apparatus, i.e., Electric Vehicle Supply Equipments~(EVSEs), is naturally increasing. Although EV charging equipment can be installed on residential premises, their absence in public spaces is often seen as a limiting factor, which restricts EV users to not travel far away from the charging station. Various governments, as well as industry players, are working to solve this issue by increasing the presence of EVSEs in public spaces. For instance, the USA, Germany, and China have allocated dedicated funds to develop the EV charging network~\cite{evs1_6} in their countries. On the other hand, companies are installing EVSEs in their parking lots for their employees~\cite{evs1_20}. Thus, we can expect major growth in publicly available EVSEs in the coming years that will reduce infrastructure availability concerns, increase users' convenience, and may further boost EV adoption.
\par
Unlike the refueling process of conventional vehicles, the charging process of EVs involves complex communication protocols and information exchange between users/EVs and EVSEs infrastructure. To initiate a charging session on a public EVSE, a user has to book a charging session, negotiate power requirements, authorize the session and payment for the service, and finally station the vehicle for the duration of the charging process. The overall charging process of EVs can be divided into two parts:~(i)~resource negotiation phase and (ii)~actual charging phase~\cite{marra2012demand, lee2019acn}.
\par
As the interactions~(between the user and EVSE infrastructure) in the former phase involve exchanging private information, such interactions are protected by the state-of-the-art communication protocols and cryptographic mechanisms~\cite{evs1_1,evs1_9}. The interactions~(between EV and EVSE infrastructure) in the latter phase primarily focus on transferring energy to recharge the vehicle and do not involve sharing of any personal information. Therefore, the signals in the charging phase are neither authenticated nor coded; these signals are exchanged in the clear. Consequently, an attacker may exploit such unprotected signals as a side channel to gain information about the EV, e.g., its battery behavior~\cite{sun2020classification}. 
\par
\textit{Motivation:} As the majority of public EVSEs are installed without proper physical access control or supervision, such equipments are accessible to anyone~\cite{evs1_3, evex5}. Thus, attackers targeting EVSE infrastructure can modify~\cite{evs1_1} EVSE's physical port and gather data related to the charging phase of benign users' EVs. In fact, recent works~\cite{sun2020classification, brighente2021tell, brighente2021evscout2} demonstrate how to use such data/signals to profile EVs with certain assumptions. Such attacks, if possible in real-world settings, can severely threaten users' privacy because attacker(s) - who have access to multiple public EVSEs - can track the movements of users who use compromised charging stations. In this paper, we investigate the extent and feasibility of such profiling of EVs in real-world scenarios. One of the major benefits of such an investigation is that it will help the community to understand the actual magnitude of EV profiling~threat. 
\par	
The key idea behind EV profiling is that each EV exhibits unique physical characteristics during a charging session. More precisely, when the State of Charge~(SoC) of the battery goes above a certain threshold (say, over 60\% or 80\%), the current and voltage drawn by the vehicle solely depend on the battery's implementation. Therefore, these physical properties - which can differ from one EV to another - can be used to create signatures of EV batteries; consequently, the signature of EVs. Authors in work~\cite{sun2020classification} demonstrate modeling the behavior of EV batteries from their charging data. Their work extracts features from analog charging signals and uses that information for battery profiling via clustering-based approach. EVScout attack~(originally EVScout1.0~\cite{brighente2021tell}, and recently EVScout2.0~\cite{brighente2021evscout2}) further improved such profiling of EVs by utilizing different machine learning~techniques. 
\par
\textit{Contributions:} In this paper, we begin with improving the state-of-the-art of EV profiling. To understand the impact of the improved EV profiling approach at scale in the real world, we emphasize on the multi-class classification~(contrary to binary classification considered in the state-of-the-art profiling approach) to evaluate its efficacy in profiling/identifying a particular EV. Furthermore, we consider datasets that vary in size, balancing, and distribution to closely simulate different settings. The major contributions of this paper are as follows:
\begin{enumerate}
	\item We propose an improved EV profiling approach that outperforms the state-of-the-art, i.e., EVScout.
	\item We exhaustively evaluate the quality of our improved approach at scale by considering a significantly large dataset of charging sessions from real EVs as well as different classification techniques, etc.
\end{enumerate} 	
\par
\textit{Organization:} The remainder of this paper is organized as follows. Section~\ref{section:background} presents a brief summary of the fundamental concepts related to our work. Section~\ref{section:threat_model} explains our threat model and attack infrastructure. Section~\ref{section:our_approach} elucidates the implementation details of our approach. Section~\ref{section:evaluation} reports our experimental evaluations. Section~\ref{section:discussion} comments on the limitations of the current practices to profile EVs. Section~\ref{section:conclusion} concludes the paper.

\section{Background}
\label{section:background}
The concept of using electric or analog data for the purpose of user profiling has been extensively studied in the literature~\cite{cronin2021charger}. The central aspect of the EV charging system is the EVSE infrastructure. A central control unit is responsible for monitoring the operation of all EVSEs connected to a particular grid. These operations include appropriate scheduling of charging processes (keeping track of power availability and maximum allowed load for the network, etc.) and constituting a gateway for secure communication between the grid and an EV (to allow user authentication, etc.). It is important to note that EVSEs are typically part of a complex network, where they can communicate with each other, an EV, or the control unit via appropriate communication interfaces. Such communications happen over a secure channel that can be wireless or wired. An EV user must be connected to the control center via a car or mobile application. The security considerations of this communication network is addressed by strong cryptographic tools and mechanisms~\cite{falk2012electric}.
\par 
The physical port on EVSEs that connects it to an EV is built upon SAE J1772 Standard~\cite{toepfer2009sae}~(cf.~\figurename~\ref{fig:system_architecture}). According to this standard, a port consists of five lead connectors. Out of these five leads, three are are connected to the grid via relays while the other two leads are used for signaling. In particular, these two leads individually carry proximity signal and pilot signal. The proximity signal verifies whether the physical connection between the EV and EVSE's port is safe and that the communication or charging can proceed. On the other hand, the pilot signal serves as a communication medium between the EV and EVSE to signal charging level, etc.
\par
The charging characteristics of the battery units used in EVs also play a part in the profiling process. Most battery units deployed in EVs today are lithium-ion batteries~\cite{cairns2010batteries}. The charging process for standard lithium-ion batteries is distinctive, where the drawn current and voltage follow a fixed profile~\cite{marra2012demand}. In particular, its charging process can be of two types, i.e., Constant Power/Constant Voltage~(CP/CV) and Constant Current/Constant Voltage~(CC/CV). In this work, we only consider the latter as sufficient data is not publicly available for CP/CV charging-based EVs. The CC/CV charging method consists of two phases:
\begin{enumerate}
\item Constant Current: It is the primary phase of charging, during which the current passed remains constant while the voltage across the battery terminals varies.
\item Constant Voltage: It is the latter phase of charging, during which the current passed drops while the voltage across the battery terminals remains constant.
\end{enumerate}
The transition from the CC to CV phase is roughly preset, but it is also ascribed by the state and condition of the EV's battery. Such transition threshold varies between 60\% and 80\% of the battery's SoC. Similar to the state-of-the-art, our approach utilizes analog signal data (e.g., current and pilot signals) obtained from the CC/CV charging phases for EV profiling. Nonetheless, our work differs in various aspects, including an improved profiling algorithm, modeling, classification approach, etc.	
\section{Threat model}
\label{section:threat_model}
EV profiling attacks~(e.g., EVScout~\cite{brighente2021evscout2, brighente2021tell}) present in the literature assume that an attacker is capable of installing a physical device - typically over EVSEs' physical port - to intercept the analog signals exchanged between EVSEs and EVs. With such a device in place, the attacker(s) can intercept, record, or transmit the observed signals to the attacker(s), where they can process the collected signals. It is worth mentioning that if such a device has wireless transmission capabilities, then tracing the original attacker(s) can become even more difficult. By tampering multiple EVSEs, the attacker(s) can have access to multiple charging sessions of different~(often, even the same) EVs. Therefore, the attacker(s) can exploit such charging data to profile the unique charging behavior of an EV's battery; which essentially means the profile of that EV.
\par
The data obtained by such a data collection practice will be unlabeled because the extracted signal is analog in nature and does not contain any personally identifying details. Manual monitoring, utilizing cameras, or collusion with local staff can make the attack sophisticated. Nevertheless, by gathering sufficient samples per EV and EVSE, the attacker(s) can identify whether a given charging session is similar to the one already present in the dataset. Therefore, the attack is able to simultaneously build training sets for multiple EVs, where the location of EVSE indicates the EV's coarse location.
\par
It is important to note that the core focus of this work is on the modeling/profiling side of the attack rather than the tampering of EVSEs for data collection.

\section{Our approach}
\label{section:our_approach}
\begin{figure}[H]
\centering
\includegraphics[trim = 0mm 155mm 0mm 15mm, clip, width=.825\linewidth]{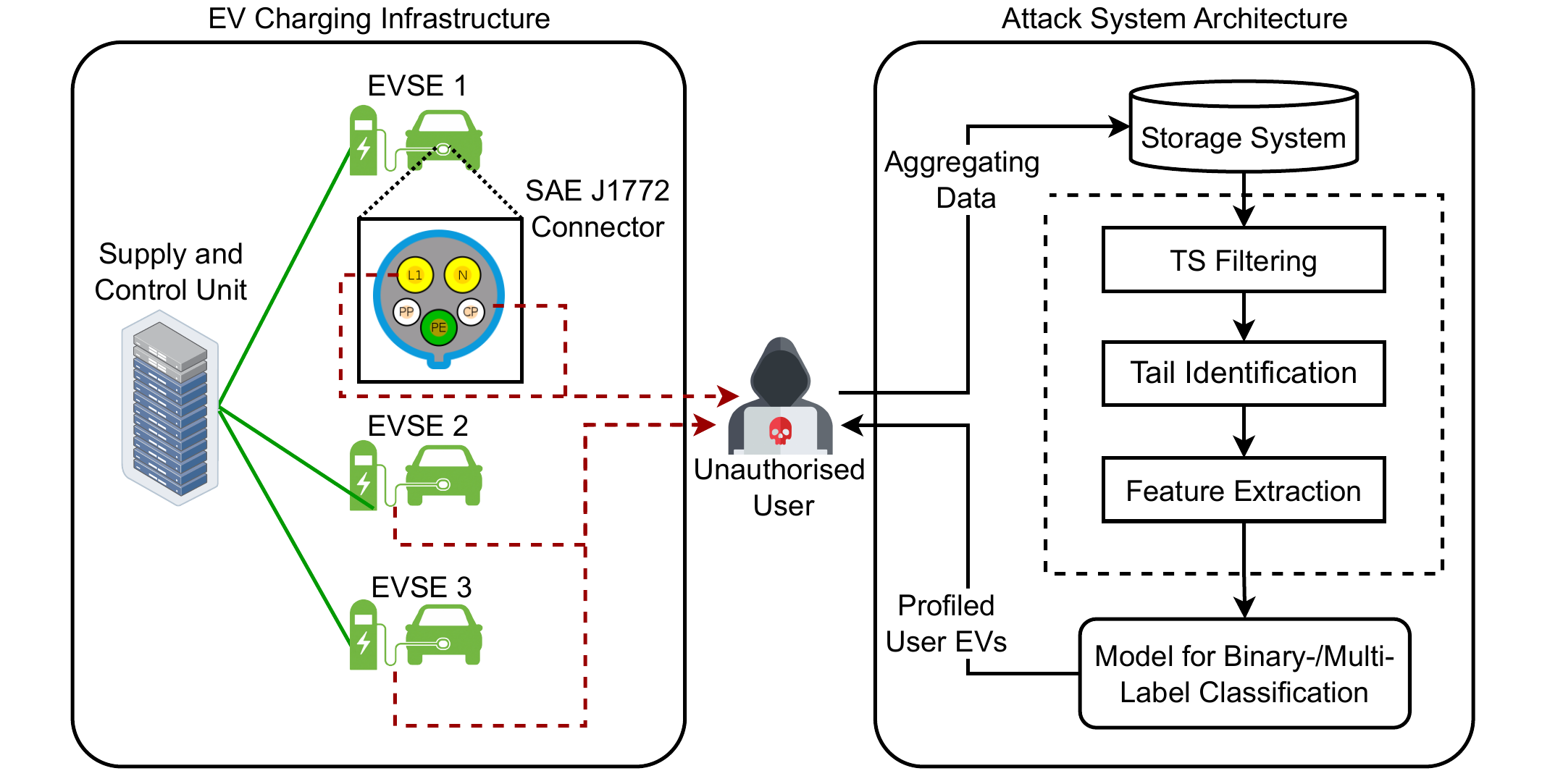}
\caption{A schematic representation of our EV profiling approach.}
\label{fig:system_architecture}
\end{figure}

\figurename~\ref{fig:system_architecture} presents an overview of our EV profiling approach. It begins with collecting charging data for EVs to build their profiles. Essentially, we want to identify characteristics that are unique to each EV. To this end, we focus on the charging behavior of EVs' batteries. Since the data available to the perpetrator is analog current-based Time Series (TS) quantities and the current passing through a battery during the CV phase varies~(cf. Section~\ref{section:background}), we leverage the variation in the current supply for profiling the battery's behavior. Following the naming system used in the work~\cite{sun2020classification}, we call the current signal TS during the CV phase, a `tail'. From a raw data sample, we must first identify its tail. Then, we extract meaningful features from the tail to build a machine learning model. It is worth mentioning that similar to current variations during the CV phase, voltage variations during the CC phase may be used for the profiling process. In the absence of such data~\cite{sun2020classification, brighente2021evscout2, brighente2021tell}, we consider only the current variations in our work.
\par
The remainder of this section elaborates on the implementation details of our improved EV profiling approach. Section~\ref{subsection:data_collection} explains the key attributes, parameters, and structure of the dataset used in our study. Section~\ref{subsection:data_preprocessing} describes our filtering process for noise elimination from TS data. Section~\ref{subsection:tail_identification} elaborates on tail identification methodology. Section~\ref{subsection:feature_extraction} covers the details of our feature extraction and classification approach. Finally, Section~\ref{subsection:algorithm} discusses the key optimizations that helped us improve our approach over the state-of-the-art of EV profiling.

\subsection{Dataset}
\label{subsection:data_collection}
\begin{figure}[H]
\centering
\includegraphics[trim = 0mm 5mm 5mm 0mm, clip, width=0.625\linewidth]{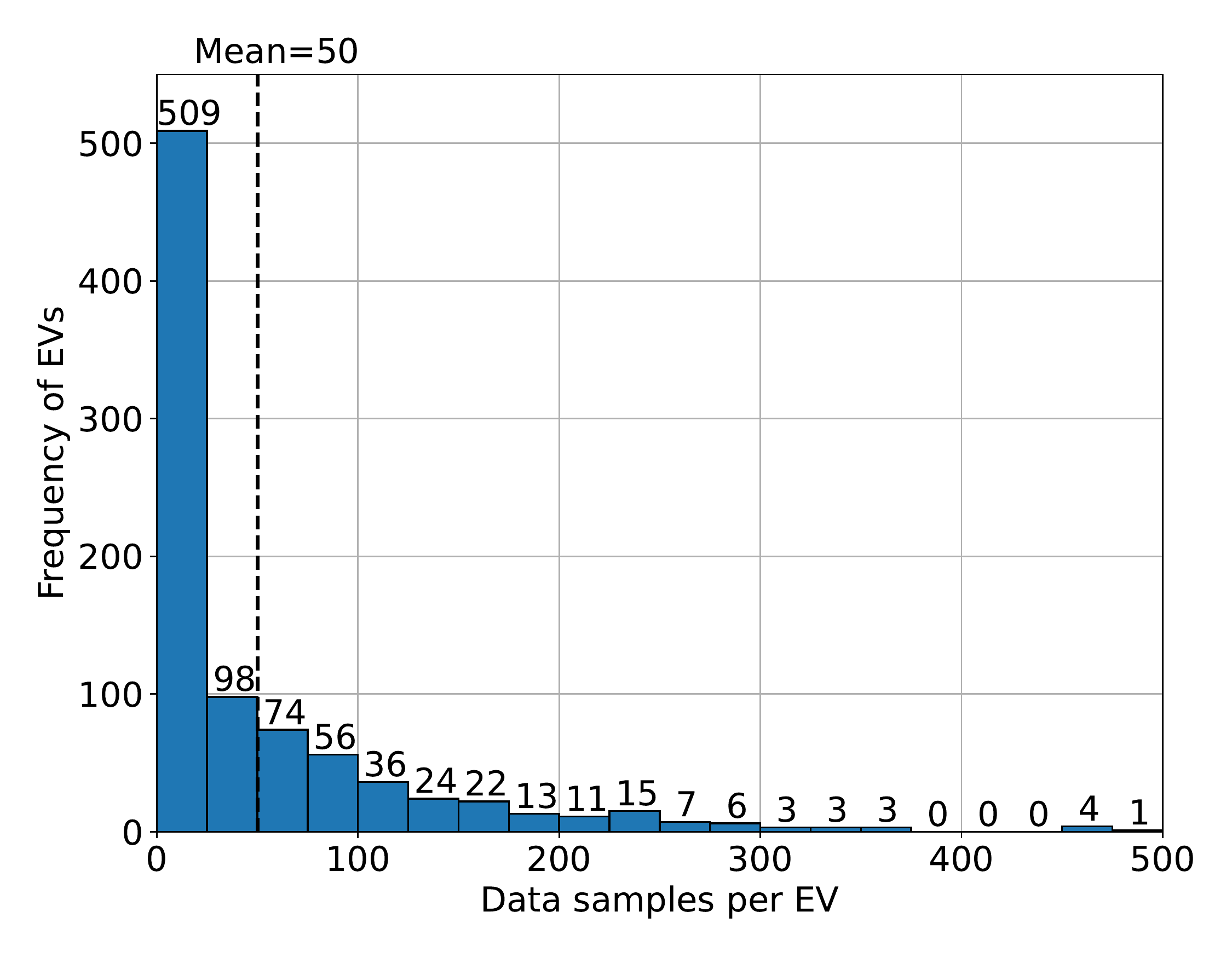}
\caption{Frequency distribution of ACN dataset.}
\label{fig:data_frequency}
\end{figure}

In this work, we use the ACN open EV charging dataset~\cite{lee2019acn}, which is the largest publicly available dataset in this category at the time of this study. It consists of data aggregated from two EVSEs that are connected to a central controller regulating power over the grid. The dataset furnishes the details of all charging sessions that had taken place over two years of span (i.e., from October 2019 to December 2021) on the CalTech campus and JPL campus EVSEs. \figurename~\ref{fig:data_frequency} shows the frequency distribution of ACN dataset. Here, over 509 unique EVs have up to 25 data samples~(i.e., charging sessions) and the average data sample per EV is 50. 
\par
The dataset contains TS data of charging current signal and pilot signal, unique user ID for each EV, and other charging parameters (e.g., connection time, total power passed, charging duration, timezone). To build our system, we only used current signal TS, pilot signal TS, and user ID. While the intercepting device (cf. Section~\ref{section:threat_model}) can only gather analog signals (i.e., current signal and pilot signal), we utilize user IDs from the dataset as labels for model training purposes only. \tablename~\ref{tab:acn_dataset} highlights the key attributes of the ACN dataset.
\begin{table}[H]
\centering
\caption{Key attributes and parameters of the ACN dataset}
\resizebox{.375\textwidth}{!}{
	\begin{tabular}{|l|c|}
		\hline
		\multicolumn{1}{|c|}{\textbf{Attribute}}                                                          & \textbf{Description}                                                \\ \hline
		Number of unique EVs                                                                    & 885                                                                 \\ \hline
		\begin{tabular}[c]{@{}l@{}}Number of data samples\\ across all EVs\end{tabular}         & 44250                                                               \\ \hline
		\begin{tabular}[c]{@{}l@{}}Average number of data\\ samples per EV\end{tabular}        & 50                                                                  \\ \hline
		\begin{tabular}[c]{@{}l@{}}EVSEs locations considered\\ in data collection\end{tabular} & \begin{tabular}[c]{@{}c@{}}2\\ (Caltech \& JPL Campus)\end{tabular} \\ \hline
		\begin{tabular}[c]{@{}l@{}}Number of common EVs\\ across locations\end{tabular}         & 81                                                                  \\ \hline
		Key parameters                                                                          & \textit{userID, pilotSignal, chargingCurrent}                                \\ \hline
	\end{tabular}
}
\label{tab:acn_dataset}
\end{table}
\par 
Due to the upper power limit of the grid, premature departure of the user, etc., it is possible that the battery of an EV is not fully charged when it disconnects from the EVSE. For these possible reasons, we observe that the TS data retrieved from the dataset varies in length. EVSE scheduling algorithms, voltage fluctuations, loose physical connection between terminals, random noise, etc. may also cause arbitrary spikes and depressions in the collected data.
\par 
At the time of retrieving the data from the dataset using its API calls~\cite{lee2019acn}, we implement a primary filter check to consider only those charging sessions that include reasonable length TS~(i.e., at least 100 data points) of both current and pilot signals as well as an identifiable user ID tag. We also excluded those EVs that have a small number of charging sessions~(i.e., less than 10) as they can not be directly used for our classification tasks. After applying further constrains~(described in Section~\ref{subsection:data_preprocessing} and Section~\ref{subsection:tail_identification}), our final dataset contains 25032 charging data samples from 530 unique EVs.	

\subsection{TS filtering}
\label{subsection:data_preprocessing}
\begin{figure}[!htbp]
	\centering
	\subfigure[Raw current signal TS.]{	    
		\includegraphics[trim = 2mm 2mm 5mm 8mm, clip, width=0.46\linewidth]{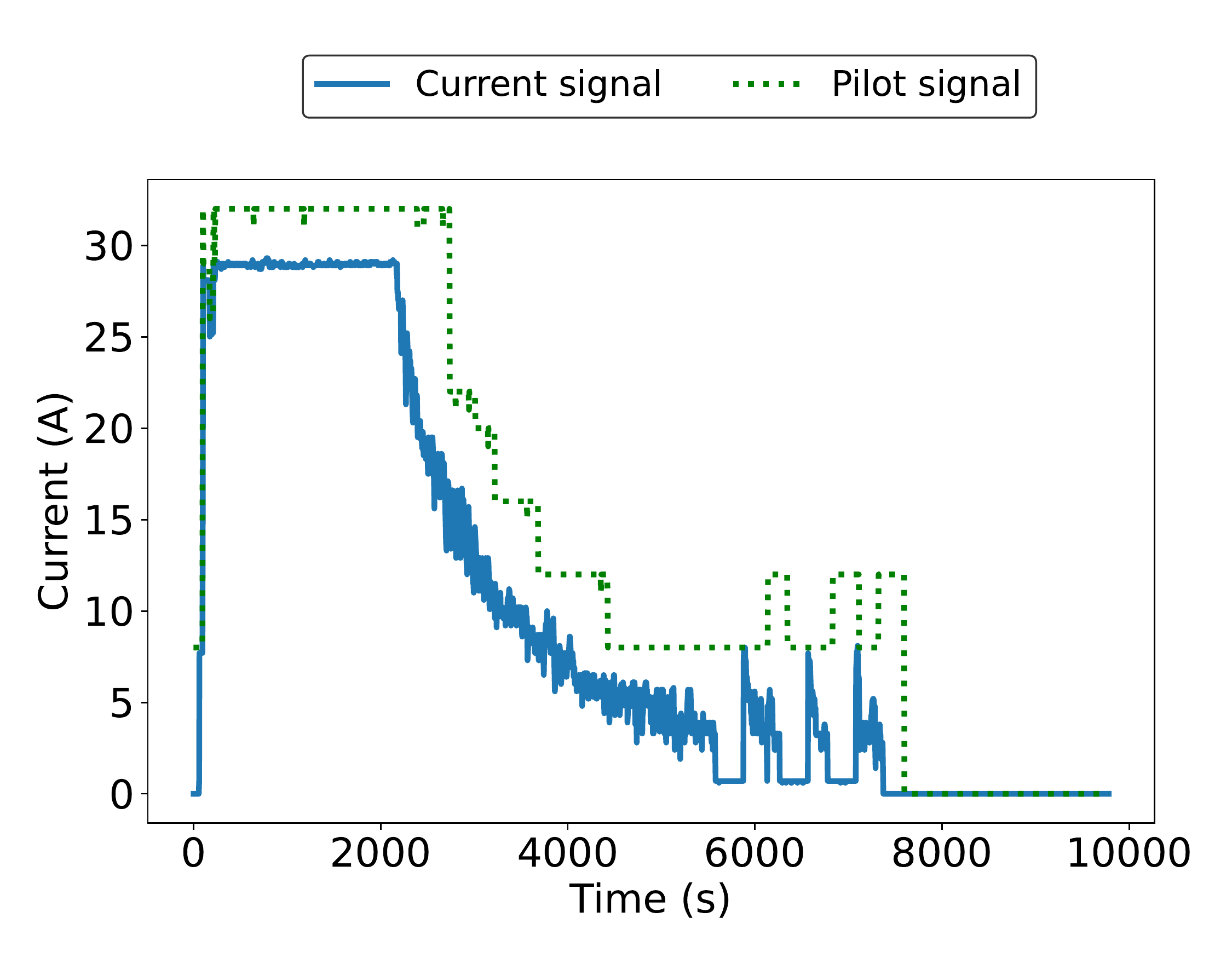}
		\label{fig:acn_datapoint}}
	\subfigure[Filtered current signal TS.]{	    
		\includegraphics[trim = 2mm 2mm 5mm 8mm, clip, width=0.46\linewidth]{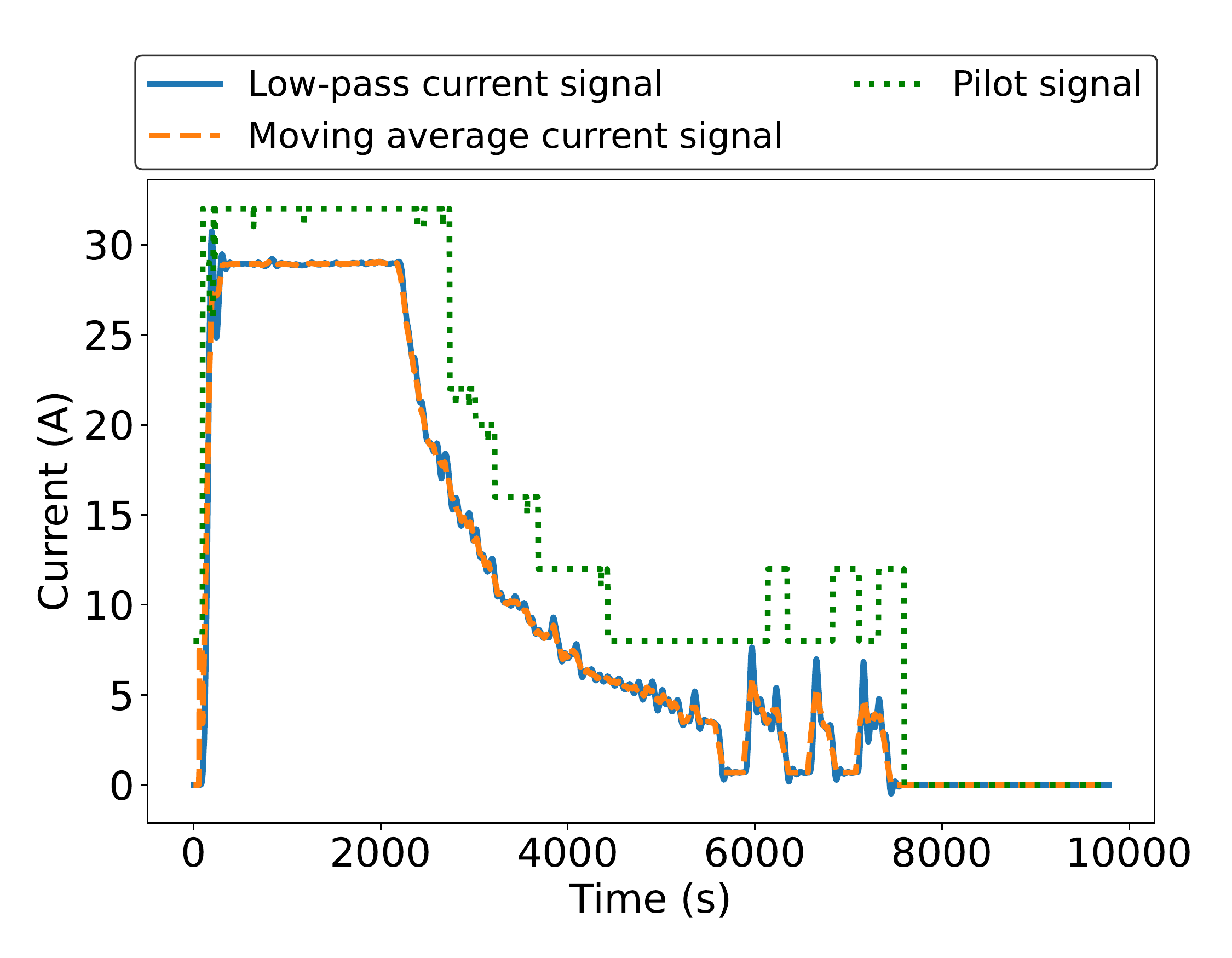}
		\label{fig:filtered_datapoint}}
	\caption{The effect of filtering on a sample current signal TS; shown along with its corresponding pilot signal TS.}
\end{figure}

In order to identify the tail in a data sample from an EV's charging session, we must appropriately segregate the CC and CV phase. \figurename~\ref{fig:acn_datapoint} shows a sample current signal TS along with its pilot signal TS. Here, the current signal remains constant during the first phase (i.e., the CC phase) and drops to zero in the second phase (i.e., dropping tail in the CV phase). Since the collected analog signal can also contain noise, it can affect the identified boundary between the CC and CV phases. Thus, we propose to filter such noise present in the samples first. By normalizing the noise, we make the two phases more distinguishable and amplify the trends in the tail; which subsequently improves the tail identification phase. To this end, we considered a variety of filters. We find that a low pass filter~\cite{karki2000active} and a moving average filter produce consistent results. \figurename~\ref{fig:filtered_datapoint} demonstrates the effect of these filters on the sample current signal TS.
\par
For the low pass filter, we can empirically find the thresholds to eliminate noise from the TS. Here, any data points beyond the thresholds are attenuated, allowing us to fine-tune the frequency of acceptable oscillations. However, attenuation of higher amplitude signals is undesirable for us because of the ringing effects. The sinusoidal spikes in the tail~(cf. \figurename~\ref{fig:filtered_datapoint}) can be relevant traits for classification. We find that such threshold-based elimination of the noise can affect the number of sinusoidal spikes, affecting the traits in the tail. As a result, the moving average filter is preferred in our system as it is able to eliminate the noise while considering average values over a localized region. For a current signal~$c$ that varies with time~$t$, we generate a moving average filtered output signal~$y(t)$ over a sliding window of size~$N$ as shown in Eq.~\ref{eq:fraction}.
\begin{equation}
\label{eq:fraction}
y(t) = \frac{1}{N} \sum_{j = - \frac{N}{2}}^{\frac{N}{2}}c(t + j).
\end{equation} 
\par
To better capture the specific characteristics of an EV's battery, we also focus on the CC phase of charging sessions. In the CC phase, we define a delta TS that is obtained by computing the difference between the pilot and current signals. The pilot signal indicates the immediate state of the EVSE. With the delta TS, we are able to gauge how effectively the EV's battery can input current with respect to the ideal peak value over the entire CC phase. Since the values in the CC phase are rather constant, the slope of descent here is less steep than in the tail. Thus, we use a moving median filter for the delta TS to remove noise and outliers. For pilot signal~$p$ and current signal~$c$ that vary with time~$t$, we generate the output delta signal~$d(t)$ over a sliding window of size~$N$ as shown in Eq.~\ref{eq:median}. 
\begin{equation}
\label{eq:median}
d(t) = p(t) - median(\{ c[ t - \frac{N}{2}, t + \frac{N}{2}] \}).
\end{equation}

\subsection{Tail identification}
\label{subsection:tail_identification}
After the TS are processed and filtered, the next step is to identify the tail in the data sample. Essentially, the tail identification algorithm should first determine the presence of the tail (i.e., the CV phase)~\cite{sun2020classification}. Thus, the algorithm attempts to find a sequence of zero-values from the end (i.e., in the reverse direction of TS data). In this reverse direction of data processing, sometime a shorter zero-value sequence may appear before the longest zero-value sequence. Such instance primarily represents a random noise spike between the two. Thus, such portions of the data sample are excluded from further processing. Once a steady zero-value state~($t_{s}$) is identified, the algorithm checks the tail to ensure that the slope of values is non-decreasing over discrete time intervals. With respect to such a strictly monotonic function of slope, we must furnish some tolerance to account for arbitrary fluctuations. Thus, we consider the tolerance function shown in Eq.~\ref{eq:tolerance}.
\begin{equation} 
\label{eq:tolerance}
y(t+1) - y(t) < \epsilon.
\end{equation}
Here, $\epsilon$ is a fixed small value. In addition to tolerating local fluctuations, we must also handle steady current spikes~(cf. three spikes towards the end of the tail in~\figurename~\ref{fig:acn_datapoint}) that may be present in the tail. Such spikes can prematurely terminate the tail extraction algorithm. Hence, we define~$T_{max}$ as the number of consecutive non-increasing values to be tolerated; any fluctuations in between resets~$T_{max}$ counter. In other words, the algorithm will continue as long as it does not encounter $T_{max}$ number of consecutive non-increasing values or the other end of the TS has arrived.
\par
In addition to the checks described above, our system employs additional validation constraints that must be met by both the extracted tail and delta TS. These constraints are:
\begin{enumerate}
\item If the extracted tail or delta series from a given sample is too small, such samples are discarded. Because a TS that is too small may not provide relevant information, instead it is prone to add irrelevant information to classifier.
\item If the extracted tail or delta series from a given sample is excessively large, such samples are discarded after manual inspection. We observed that large TS were mainly constant~(even zero) valued. Such TS only increase the execution time for feature extraction step and do not contribute any additional relevant information in it.
\item In some charging sessions, especially those that have been stopped prematurely, the tail or delta series mainly consist of zero values. Such samples with zero-valued tail or delta series are discarded as they hardly contribute any relevant information for subsequent stages of our system.
\end{enumerate}

\subsection{Feature extraction}
\label{subsection:feature_extraction}
Instead of taking a restrictive approach of manually deciding parameters or performing unsupervised learning using Random Forest, we use a feature extraction tool to get a comprehensive list of relevant features from our tail and delta TS data. The key benefit of such an approach is that the extracted features can be suitably adapted for the target classification algorithms.
\par
In particular, our feature extraction process is based on Scalable Hypothesis tests~(i.e., tsfresh)~\cite{christ2018time} library, which can be easily integrated into the python interface and scikit-learn~\cite{pedregosa2011scikit} models. Furthermore, its support for accelerating the computations using GPUs and multi-threading the operations, helps us reduce the executing time of the entire process. The tsfresh library extracts about 1500 features in total for the extracted tail and delta TS data. As these many features can result in overfitting~(particularly for smaller datasets), we fix the maximum Number of Features~($NoF$) to be used for classification. Then, we select the most relevant features using the scikit $SelectKBest$ method. The mathematical function used for such selection can be $chi2$~\cite{liu1995chi2} or $f\_classif$~\cite{st1989analysis}. $chi2$ measures the dependence between stochastic variables while $f\_classif$ is based on the ANOVA method and helps in selecting features based on their covariance.
\par
\textit{Classification:}	
Next, we build a machine learning model from the extracted features. The classifiers that we consider are Random Forest~(RF)~\cite{breiman2001random},  Decision Trees~(DT)~\cite{loh2011classification}, k-Nearest Neighbors~(kNN)~\cite{guo2003knn}, and Support Vector Machines~(SVM)~\cite{suykens1999least}. These four classifiers output the highest performance scores in the EVScout attack~\cite{brighente2021evscout2, brighente2021tell}, and thus, are used for benchmarking our approach. The classification workflow and hyper-parameters used for these classifiers are discussed in Section~\ref{subsection:evaluation_setup}. 

\subsection{Key optimizations}
\label{subsection:algorithm}
The baseline approach to profile EVs using their charging data has been set by the work~\cite{sun2020classification}. EVScout~\cite{brighente2021tell, brighente2021evscout2} improves such profiling by employing different machine learning models. Our work further improves EV profiling by defining:~(i)~a finer TS filtering procedure, particularly by using sliding window-based mean, median filters, etc.~(cf. Section~\ref{subsection:data_preprocessing}); (ii)~multi-level checks for the tail identification process, particularly by tolerating both local fluctuations and current spikes, etc.~(cf. Section~\ref{subsection:tail_identification}); and (iii)~an enhanced feature extraction method, particularly by utilizing different mathematical functions, etc.~(cf. Section~\ref{subsection:feature_extraction}). Furthermore, we also consider the following two optimizations to improve the quality of our approach.
\begin{enumerate}
\item \textit{Dataset:} 
As discussed in Section~\ref{subsection:data_collection}, we use the ACN EV charging dataset~\cite{lee2019acn} in our study. It contains the charging data for 885 unique EVs from two EVSEs at different locations. Even after excluding TS that could not satisfy our various constraints~(minimum data points per sample, minimum charging sessions per vehicle, validations on the extracted tail and delta TS, etc.), our final dataset contains charging data for 530 unique EVs. The number of unique EVs considered in our study is at least about four times~(cf. \tablename~\ref{tab:filetypes}) the number of unique EVs considered in other studies~\cite{sun2020classification, brighente2021tell, brighente2021evscout2}. The larger amount of charging data has helped us identify traits for better TS filtering, tail identification, and classification model tuning. More importantly, it enabled us to sub-sample datasets with discrete data distributions (i.e., normal, uniform) that mimic different real-world trends.
\item \textit{Q-balancing:}
In order to assess the performance of EVScout with the imbalance in data for different EVs, the authors~\cite{brighente2021tell} specify a parameter called $Q$. $Q$ is defined as the ratio of the number of data samples associated with the target EV to the number of data samples associated with all other EVs combined in the dataset. The value of $Q$ varies between [1, 5]. Although a higher value of $Q$ may reflect unbalanced settings, it heavily skews/biases the dataset towards the target EV. Even with $Q=1$, the dataset still remains skewed towards the target EV when considering multi-class classification. Thus, such a definition of $Q$ is not suitable for reflecting real-world implications. To this end, we define $Q'$ as the ratio of the number of data samples associated with all other EVs combined to the number of data samples associated with the target EV. Simulating dataset imbalance with $Q'$ will not\footnote{Increasing $Q$ values, increases data samples for the target class. However, increasing $Q'$ values, increases data samples for all other classes.} favor the target class. Thus, $Q'$ is a more suitable parameter for practical usage.
\end{enumerate}

\section{Evaluation}
\label{section:evaluation}
To evaluate the performance of our improved EV profiling approach, we first benchmark it against the state-of-the-art EV profiling attack~\cite{brighente2021evscout2, brighente2021tell}. Next, we assess the quality of our improved approach the in real-world settings, i.e., by using datasets of significantly larger sizes and different data distributions. We would like to reiterate that our ultimate goal is to understand the feasibility of such EV profiling threats in the real world. We explain our evaluation setup in Section~\ref{subsection:evaluation_setup} and discuss our evaluation results in Section~\ref{subsection:results}.
\subsection{Evaluation setup} 
\label{subsection:evaluation_setup}
We follow the standard operating procedures for machine learning classification tasks throughout our implementation and evaluations. All our experiments have been conducted over an 80\%-20\% stratified train-test split of the data. To increase the statistical significance of the results, we repeated each experiment five times using a different 80\%-20\% partition of data. We report the mean scores from these five runs unless stated otherwise. We learn the model parameters for a given classifier (i.e., RF, DT, kNN, and SVM) on the training set using grid search with 5-fold stratified cross-validation~(i.e., $GridSearchCV$). \tablename~\ref{tab:hyperparameters} lists the hyper-parameters used for the classifiers.
\begin{table}[H]
\centering
\caption{Hyper-parameters used for GridSearchCV optimization}
\resizebox{.375\textwidth}{!}{
	\begin{tabular}{|l|l|l|}
		\hline
		\multicolumn{1}{|c|}{\textbf{Classifier}}                                              & \multicolumn{1}{c|}{\textbf{Parameters}} & \multicolumn{1}{c|}{\textbf{Values}} \\ \hline
		\multirow{2}{*}{Random Forest}                                                    & n\_estimators                            & 5, 10, 15, 20, 30, 50                \\ \cline{2-3} 
		& max\_depth                               & None, 3, 5, 10, 15, 25               \\ \hline
		\multirow{3}{*}{\begin{tabular}[c]{@{}l@{}}k-Nearest\\ Neighbour\end{tabular}}    & n\_neighbors                             & 3, 5, 7, 9, 11, 13, 15               \\ \cline{2-3} 
		& metric                                   & euclidean, manhattan, cosine         \\ \cline{2-3} 
		& weights                                  & uniform, distance                    \\ \hline
		\multirow{2}{*}{Decision Tree}                                                    & criterion                                & gini, entropy                        \\ \cline{2-3} 
		& max\_depth                               & None, 6, 10, 18                      \\ \hline
		\multirow{3}{*}{\begin{tabular}[c]{@{}l@{}}Support Vector\\ Machine\end{tabular}} & kernel                                   & poly, rbf, sigmoid                   \\ \cline{2-3} 
		& regularization                           & 1, 10, 100, 1000                     \\ \cline{2-3} 
		& gamma                                    & 1$\mathrm{e}$-4, 1$\mathrm{e}$-3                           \\ \hline
	\end{tabular}
}
\label{tab:hyperparameters}
\end{table}
\par
\ankit{As explained in Appendix~\ref{appendix:NoF}, we empirically chose $NoF=100$ for all binary classification tasks and  $NoF=200$ for all multi-class classification tasks.} We would also like to highlight that $SelectKBest$ method~(cf. Section~\ref{subsection:feature_extraction}) is executed for training set only. Since $chi2$ function can operate only between the values~$[0,1]$, a $MinMaxScaler$ is used for transforming the train and test set, separately. All experiments have been evaluated using standard classification metrics. As a standard practice, we~report the performance using the accuracy metric, where the true positives and true negatives are crucial. We use F1-score, where the false negatives and false positives are more~important.

\subsection{Results}
\label{subsection:results}
We designed different experiments to thoroughly evaluate the quality of our improved approach. We first benchmark our approach against EVScout in Section~\ref{section:resultsQBal}. We assess the quality of our improved approach with multi-class classification in Section~\ref{section:resultsMCClass}. In Section~\ref{section:resultsSubSample}, we evaluate the performance of our approach over sub-sampled datasets that have different synthetic data distributions.
\subsubsection{Binary classification}
\label{section:resultsQBal}
\begin{figure}[H]
	\vspace{-1.25em}
\centering
\subfigure{
	\centering
	\includegraphics[trim = 0mm 136mm 0mm 0mm, clip, width=0.555\linewidth]{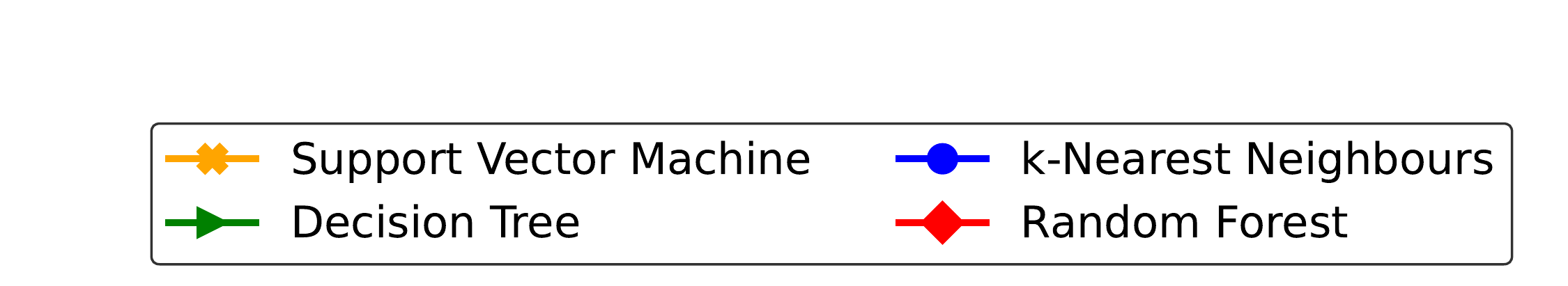}
}
\addtocounter{subfigure}{-1}
\vspace{-0.8em}

\subfigure[EVScout1.0 (Left) with $Q$-balance and our work (Right) with $Q'$-balance. \vspace{-2em}]{	    
	\includegraphics[trim = 0mm 0mm 0mm 45mm, clip, width=0.49\linewidth]{./images/graphs/EVScout1_QBalance_graph.pdf}
	\includegraphics[trim = 0mm 0mm 0mm 45mm, clip, width=0.49\linewidth]{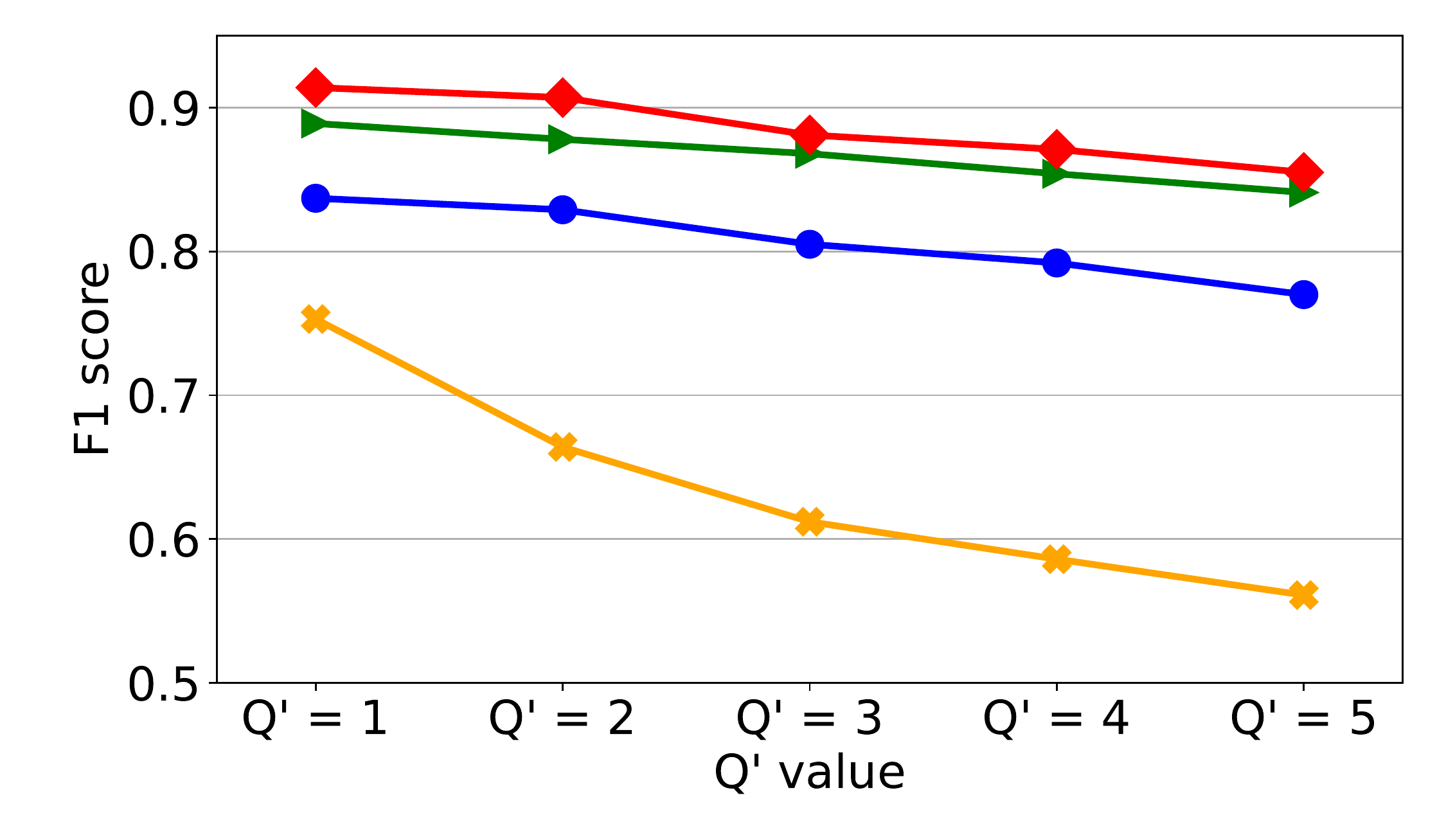}
	\label{fig:evscout1_qbalancing}}

\subfigure{
	\centering
	\includegraphics[trim = 0mm 136mm 0mm 0mm, clip, width=0.55\linewidth]{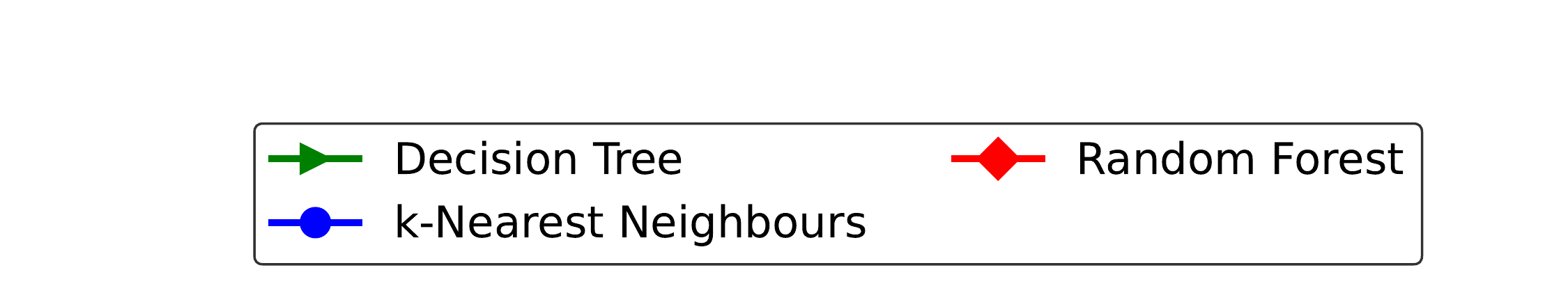}
}
\addtocounter{subfigure}{-1}
\vspace{-0.8em}

\subfigure[EVScout2.0 (Left) with $Q$-balance and our work (Right) with $Q'$-balance.]{	    
	\includegraphics[trim = 0mm 0mm 0mm 45mm, clip, width=0.49\linewidth]{./images/graphs/EVScout2_QBalance_graph.pdf}
	\includegraphics[trim = 0mm 0mm 0mm 45mm, clip, width=0.49\linewidth]{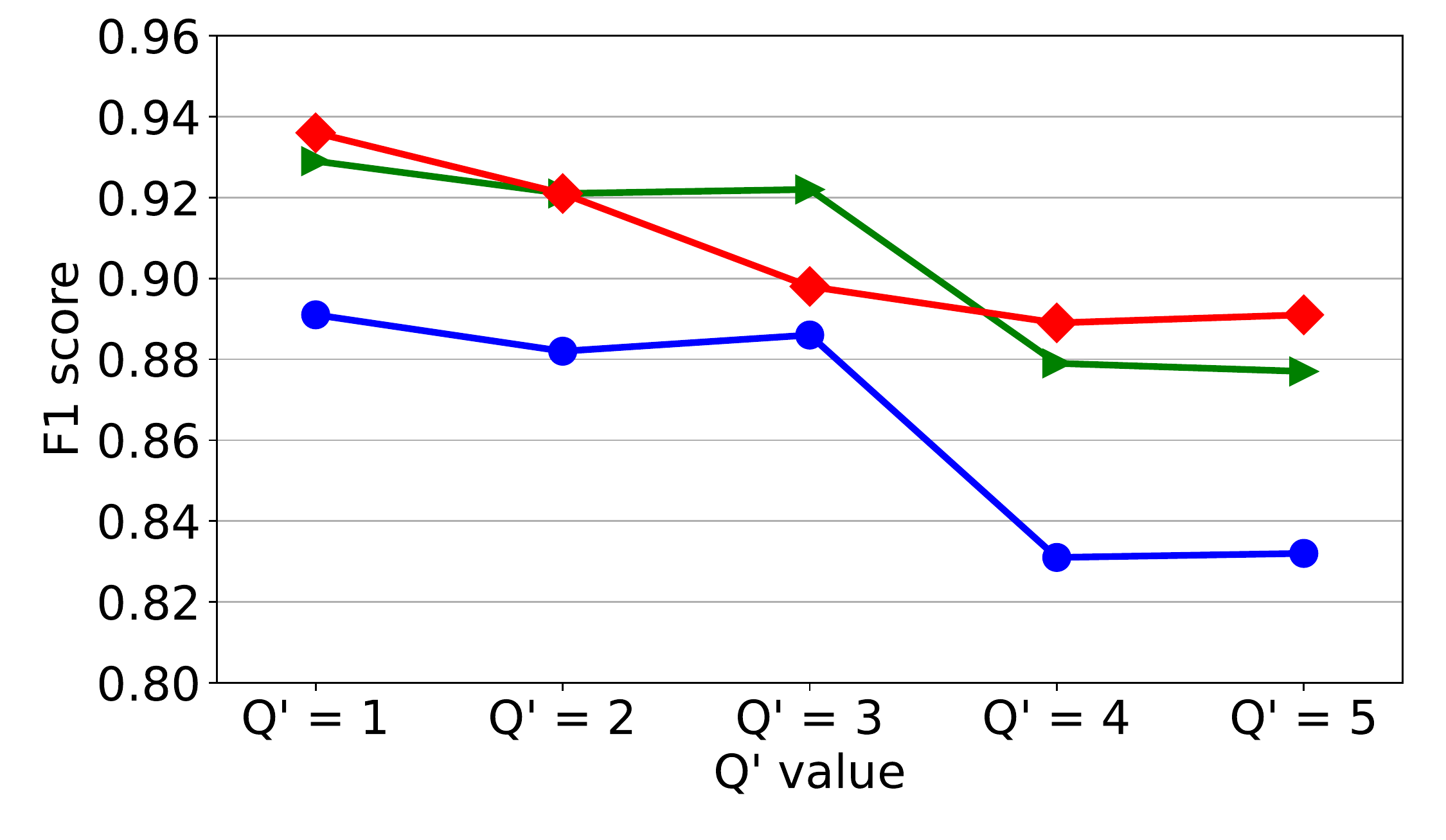}
	\label{fig:evscout2_qbalancing}}
\caption{F1-scores for binary classification.}
\label{fig:binary}
\end{figure}

EVScout uses binary classification with different values of $Q$ to evaluate the quality of EV profiling. Therefore, we use the same method of creating $N$ binary classifiers~(i.e., One-vs-All strategy for each qualifying EV) as a baseline to compare our improved approach with EVScout. For a fair comparison, we use the original $Q$-balancing with EVScout implementations and our $Q'$-balancing for our implementation. To vary the values of both $Q$ and $Q'$ between [1, 5], we keep the number of data samples associated with a given target EV constant and change the number of data samples associated with all other EVs accordingly~(cf. Section~\ref{subsection:algorithm}). In particular, we consider EVs with at least 50 data samples to satisfy different values of $Q$ and $Q'$ over each such EV. As far as the number of EVs is considered, we use the same-sized dataset as EVScout implementations, i.e., 25 EVs for EVScout1.0 and 140 EVs for EVScout2.0. \figurename~\ref{fig:binary} shows the overall average F1-scores across all $N$ binary classifiers. \figurename~\ref{fig:evscout1_qbalancing} depicts that the SVM classifier performs the worst among all four classifiers on either approach. Henceforth, we exclude the SVM classifier from subsequent experiments due to its overall lower performance.
\par
\figurename~\ref{fig:evscout1_qbalancing} and \figurename~\ref{fig:evscout2_qbalancing} show that our approach outperforms both EVScout1.0 and EVScout2.0 in terms of achieving better classification scores across different values of $Q$ and $Q'$. In fact, the difference in F1-scores of EVScout implementations and our approach substantially increases with an increasing value of $Q$ and $Q'$. It is important to note that our system yields more stable/consistent results over different values of $Q'$, i.e., the drop in our system's performance is not as steep as EVScout implementations. Therefore, we can conclude that our approach significantly improves the state of the art and establishes new benchmarks for EV profiling via charging~data.

\subsubsection{Multi-class classification}
\label{section:resultsMCClass}
The binary classification approach effectively uses $N$ binary classifiers for $N$ EVs, where each classifier identifies whether a given test sample belongs to its corresponding EV or not. In practice, the outputs of all such classifiers are aggregated/ranked using suitable scoring methods to find the probabilistic best match for a given test sample. Such an approach is generally adequate to tailor classification of specific target classes and may not scale with an increasing number of classes. Thus, we advocate for a multi-class classification - that is indeed designed to handle multiple classes - for profiling multiple EVs at scale .
\par
To test the general scalability of our improved EV profiling approach, we create different-sized datasets for multi-class classification. \tablename~\ref{tab:filetypes} presents a summary of the four datasets~(namely, $Small$, $Medium$, $Large$, and $Complete$) used in our evaluations. The $Complete$ dataset is our final filtered ACN dataset from which the other three datasets are generated in a stratified manner. The $Small$ and $Large$ datasets mimic the number of EV classes used in EVScout1.0 and EVScout2.0. Since multiple combinations can be generated for $Small$, $Medium$, and $Large$ datasets, we report their respective mean results.

\begin{figure}[!htbp]
\centering
\includegraphics[trim = 2mm 2mm 5mm 10mm, clip, width=0.51\linewidth]{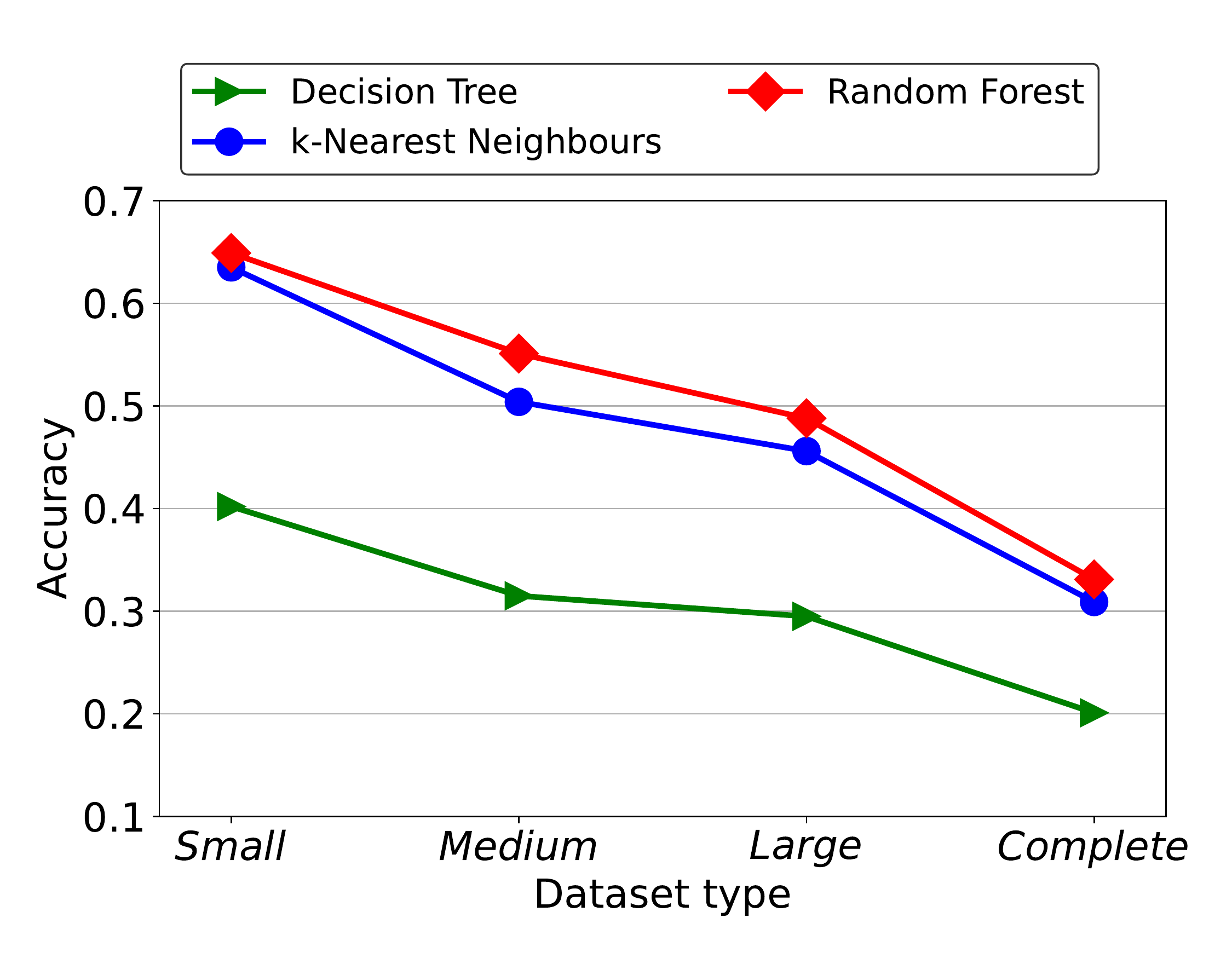}
\caption{Accuracy scores of the improved EV profiling approach for multi-class classification over different-sized datasets.}
\label{fig:multiclass_classification}
\end{figure}

\begin{table}[!htbp]
\centering
\caption{Different datasets used for multi-class classification.}
\resizebox{.35\textwidth}{!}{
	\begin{tabular}{|c|c|c|}
		\hline
		\textbf{Dataset size} & \textbf{\begin{tabular}[c]{@{}c@{}}Number of\\unique EVs\end{tabular}}
		& \textbf{Dataset used in} \\ \hline
		$Small$                 & 25                            &       \begin{tabular}[c]{@{}c@{}} EVScout1.0~\cite{brighente2021tell};\\Covers 16 EVs of~\cite{sun2020classification}\end{tabular}  \\ \hline
		$Medium$                & 75                            & -                        \\ \hline
		$Large$                 & 140                           & EVScout2.0~\cite{brighente2021evscout2}               \\ \hline
		$Complete$              & 530                           & Our dataset          \\ \hline
	\end{tabular}
}
\label{tab:filetypes}
\end{table}
\figurename~\ref{fig:multiclass_classification} reports the accuracy scores of our improved approach for multi-class classification over $Small$, $Medium$, $Large$, and $Complete$ datasets. Our results show that the system's performance over $Small$ dataset moderately decreases compared to binary classification. Such degrade in performance can be attributed to the fact that binary classification uses One-vs-All strategy, where a test sample is tried against all $N$ classifiers and \ankit{the probabilistic final label corresponds to the maximum score over $N$ classifiers. On the other side, a sample in multi-class classification can be assigned to one and only one label.} Nevertheless, the performance of the system degrades linearly with the increasing size of the dataset tested, irrespective of the classifier used.

\par
\begin{figure}[H]
\centering
\includegraphics[trim = 0mm 0mm 0mm 0mm, clip, width=0.51\linewidth]{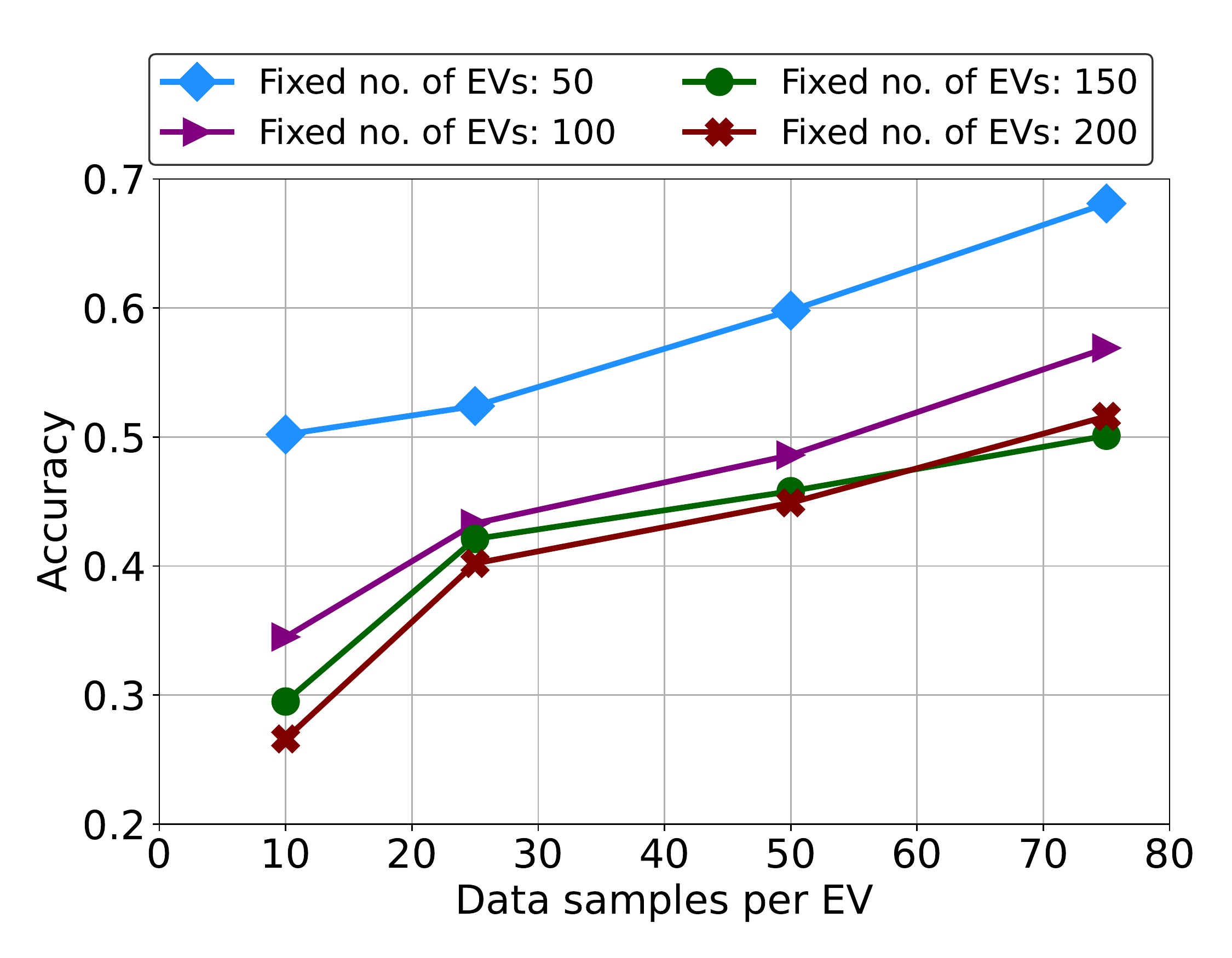}
\caption{Accuracy scores of the improved EV profiling approach for multi-class RF classification over fixed number of EVs as well as fixed number of data samples per EV.}
\label{fig:fixed_data}
\end{figure}

The degradation in performance can also partly be attributed to the deviation in the number of data samples per EV in the dataset. By regulating the number of data samples per EV, we can expect an increase in the performance. To further investigate the issue, we set up a new experiment, where both the number of EVs and the number of data samples per EV are fixed for multi-class RF classification. RF is chosen here because it was the top performer among the three classifiers tested above. \figurename~\ref{fig:fixed_data} reports the accuracy scores of our system for such setups (i.e., 50, 100, 150, and 200 EVs; 10, 25, 50, and 75 data samples per EV).
Our results shown in \figurename~\ref{fig:fixed_data} confirm our speculations, i.e., the performance of our system  increases with an increasing number of data samples per EV for a given set of EVs. However, increasing the number of EVs in the dataset negatively affects classification quality. Therefore, the performance of the system can be characterized as a trade-off between the number of EVs and the number of data samples per EV considered in the classification task. \ankit{To summarize, our results suggest that even our improved charging data-based EV profiling approach may not reliably identify a high number of EV classes.}

\subsubsection{Sub-sampled datasets with synthetic distributions}
\label{section:resultsSubSample}
As shown in \figurename~\ref{fig:data_frequency}, the original ACN dataset has an asymmetrical data distribution. As a result, when such a dataset is randomly sampled for multi-class classification, the tested system may not perform optimally. In particular, the system may not effectively train for EV classes having a smaller number of data samples, resulting in the overall lower performance of the system. We strictly filtered the original ACN dataset with several constraints~(minimum data points per sample, minimum charging sessions per vehicle, validations on the extracted tail and delta TS, etc.; cf. Section~\ref{subsection:data_collection}) to minimize such possibilities in our $Complete$ dataset.
\par
\begin{figure}[H]
\centering
\includegraphics[trim = 5mm 5mm 5mm 5mm, clip, width=0.51\linewidth]{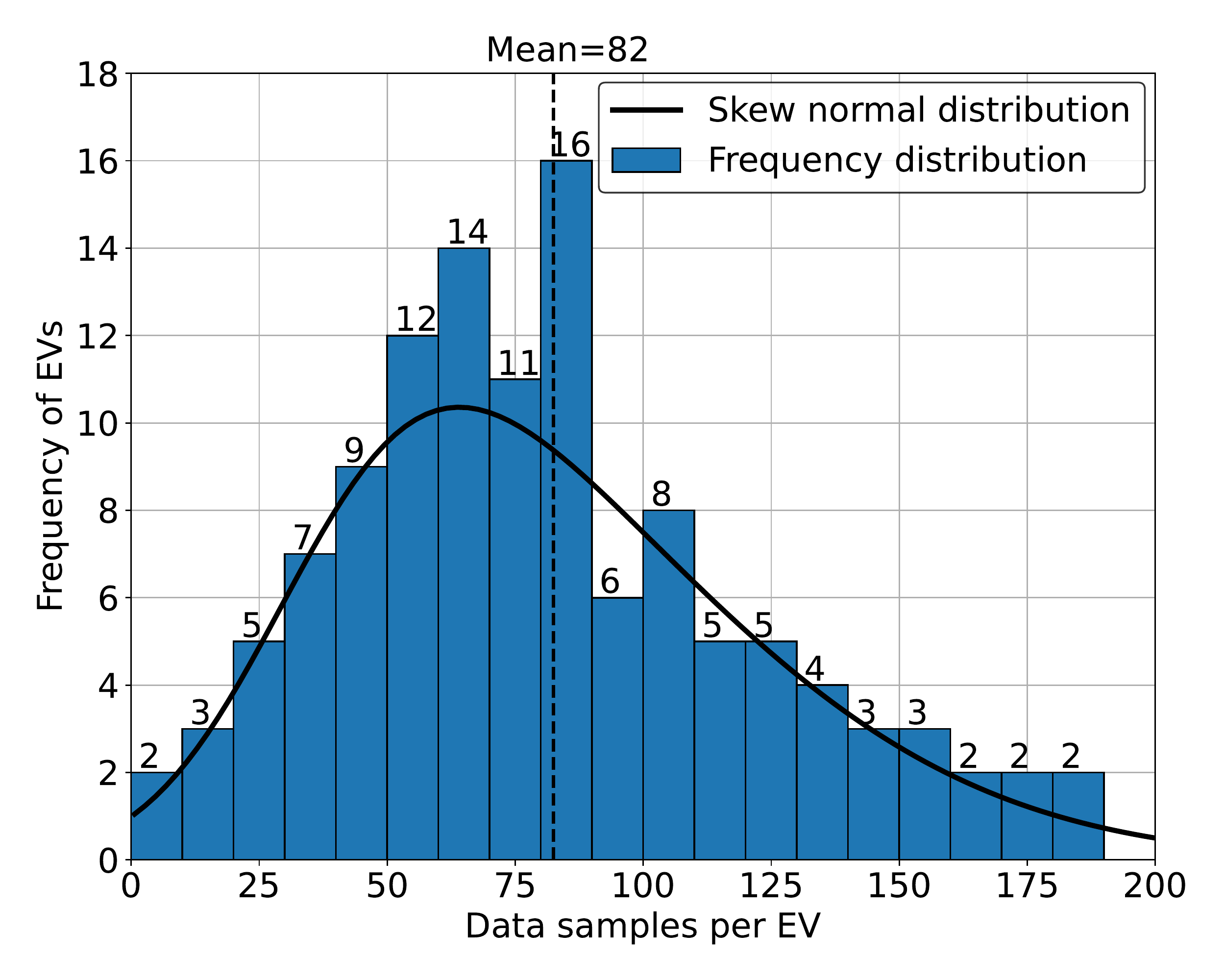}
\caption{Synthetic normal distribution generated from $Complete$ dataset.}
\label{fig:normal_frequency}
\end{figure}

To further investigate the feasibility of our improved EV profiling approach to profile multiple EVs in the real world with optimal charging data, we sub-sample our $Complete$ dataset to synthetically generate datasets with normal and uniform distributions. \figurename~\ref{fig:normal_frequency} shows the frequency distribution of our quasi-normal dataset, with the bell curve peaking near the mean for 119 EVs. On the other hand, our uniform dataset contains 6 EVs per bin. The number of EVs in these two datasets is comparable to our $Large$ dataset.
\par
\begin{figure}[H]
\centering
\includegraphics[trim = 5mm 5mm 5mm 5mm, clip, width=0.51\linewidth]{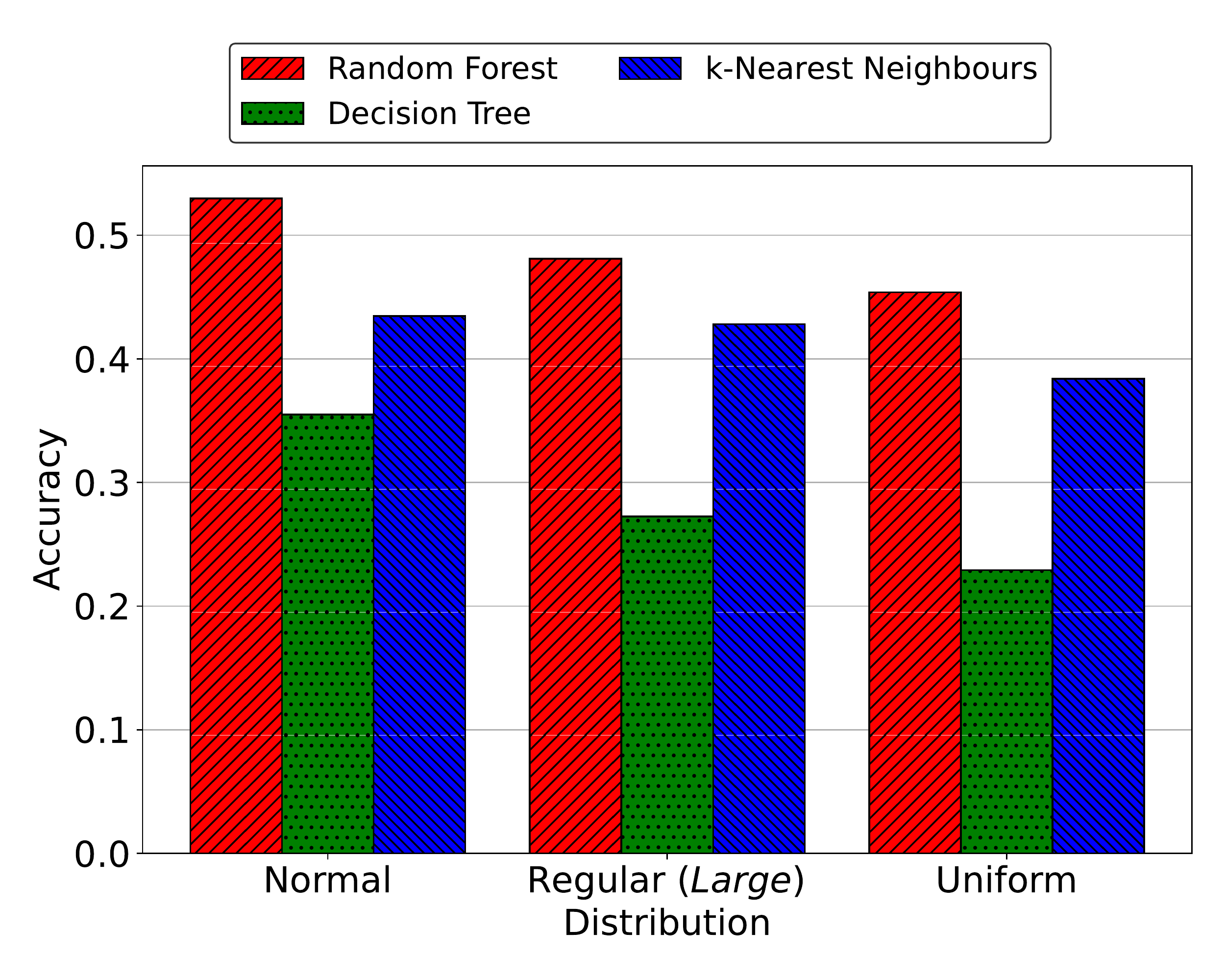}
\caption{Accuracy scores of the improved EV profiling approach for multi-class classification over datasets with different data distribution.}
\label{fig:distribution_score}
\end{figure}

\figurename~\ref{fig:distribution_score} presents the accuracy scores of our EV profiling approach for multi-class classification over datasets with regular~(i.e., $Large$), normal, and uniform distributions. When compared to the regular dataset, our system performs slightly better on the normal dataset and slightly worse on the uniform dataset. Such behavior concurs with our findings in~\figurename~\ref{fig:fixed_data}, i.e., the system's performance increases with decreasing number of EVs as well as with an increasing number of data samples per EV. Both the normal and uniform datasets have lesser EVs than the regular dataset. Hence, the number of data samples per EV is the decisive factor here. Specifically, the normally distributed dataset encourages EV classes with a high number of data samples while the uniformly distributed dataset trims such classes; thus, the difference in performance.
\par
Normal distributions hold statistical importance as they are often used to represent real-world random variables having unknown distributions. In our case, the system’s performance over the normally distributed dataset is roughly equivalent to flipping a coin, i.e., a 50\% chance of being accurate. \ankit{Therefore, we argue that even our improved charging data-based EV profiling approach may not be suitable to profile and individually identify the increasing number of EVs in practice.}

\section{Limitations}
\label{section:discussion}
In this work, we assess the feasibility of EV profiling threats in identifying individual EVs at a larger scale. In our thorough evaluations, we consider multiple classification techniques, simulate differently distributed datasets, etc. We believe the following two aspects can be perceived as potential limitations of our work.

\subsection{Dataset}
The ACN dataset used in our study is the largest publicly available EV charging dataset. Our dataset preprocessing and stringent filtering mechanisms yield a dataset that contains charging data of 530 unique EVs. As shown in~\figurename~\ref{fig:data_frequency}, the original ACN dataset has a low number of EVs that have a high number of data samples (say, over 150). Thus, our final dataset also inherits such an imbalance. We employ $Q'$-balancing to handle the issue. However, one may argue that the implications of EV profiling threats~(especially, for model training) can only be gauged in detail by using even larger datasets. \ankit{Nonetheless, our dataset is at the least about four times the dataset used in any other study in the literature~\cite{sun2020classification, brighente2021tell, brighente2021evscout2}.} Another related issue is the application of recently emerging deep learning methods. In the absence of sufficiently large datasets, such methods have not been used due to the risk of overfitting.

\subsection{Analog data}	
Another ubiquitous limitation of the current EV profiling approaches is the use of analog data itself. The presence of suitable user/EV identifiers in the public datasets has enabled the application of supervised learning. As explained in Section~\ref{section:threat_model}, the analog data intercepted by the tampering device does not contain any personally identifying details. In the absence of such labels, the task of profiling individual EVs may become even more difficult. Finally, the impact of EVSE architectures (e.g., scheduling behavior) and modern battery charging techniques (e.g., fast charging) on the intercepted analog signals needs further exploration.

\section{Conclusion}
\label{section:conclusion}
EVs are perceived as the long-term ecological alternative to conventional fuel-powered vehicles. With several worldwide awareness movements running to encourage the adoption of sustainable energy, EVs are becoming increasingly popular. In fact, several countries have already started to substantially develop their infrastructure to increase the use of EVs.
\par
While the foundation of EV transport is being laid, privacy remains one of the critical concerns for its potential users. Recent works have demonstrated a possibility of identifying EVs using the analog electrical data exchanged during the EV charging process. Thus, it becomes crucial to investigate the feasibility and magnitude of such profiling threats at scale. In this work, we propose an improved EV profiling approach that outperforms the state-of-the-art. We evaluate its performance, through a series of experiments, to profile EVs in the real world. \ankit{Our results show that even with our improved approach, profiling and individually identifying the increasing number of EVs appear extremely difficult in practice.} In the future, we will investigate the existence of other avenues that may lead to practical EV profiling at scale. If such avenues exist, we will work towards mending such possibilities.

\balance
\bibliographystyle{IEEEtranS}
\bibliography{bib}
\balance

	\appendices
	\counterwithin{table}{section}
	\counterwithin{figure}{section}
	\renewcommand{\thesection}{\Alph{section}}	
	\section{$NoF$ variation}
	\label{appendix:NoF}
\begin{figure}[H]
		\centering
		\subfigure[Binary classification]{	    
			\includegraphics[trim = 0mm 0mm 7mm 10mm, clip, width=0.6\linewidth]{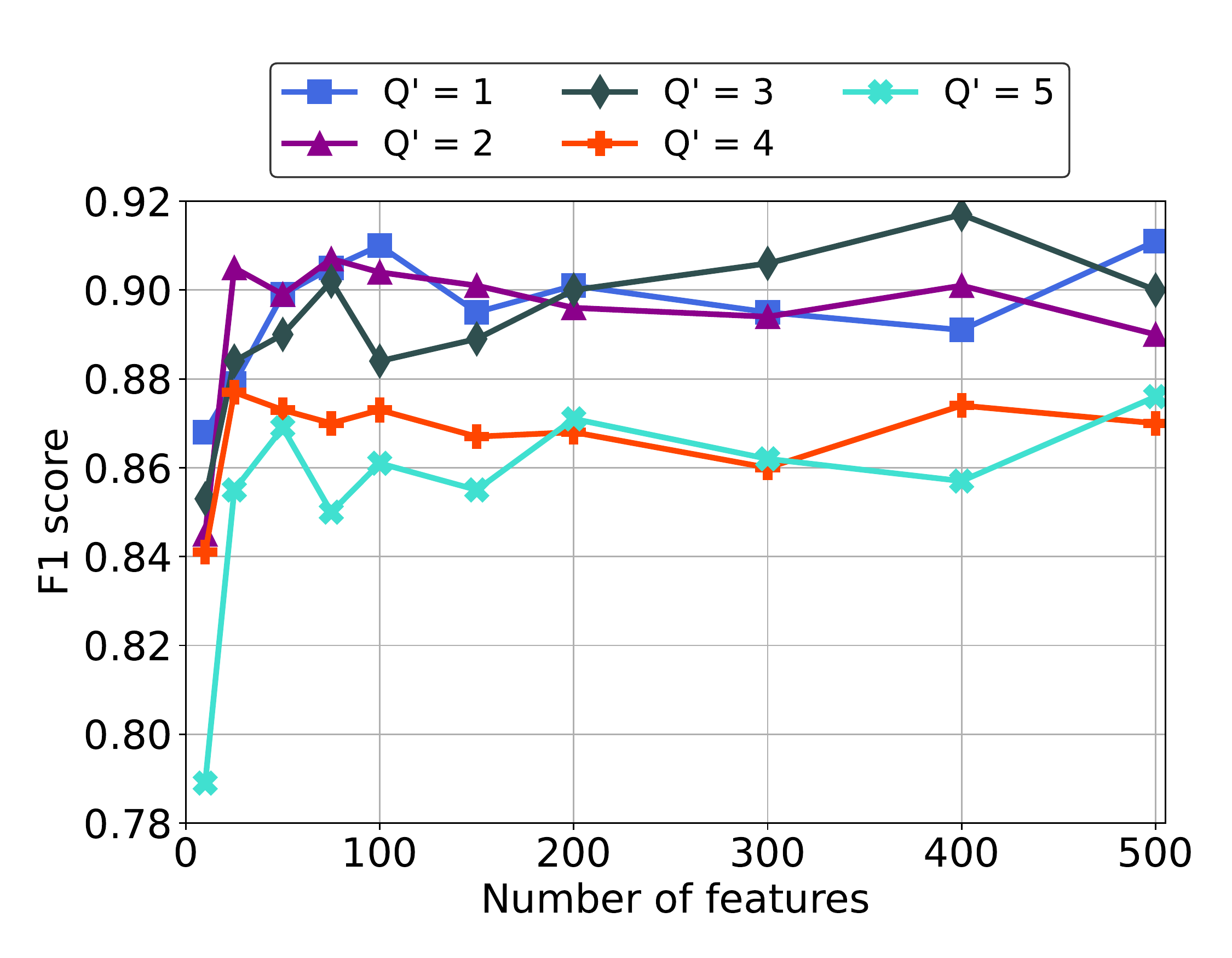}
			\label{fig:nof_qbalance}}
		
		\subfigure[Multi-class classification]{	    
			\includegraphics[trim = 0mm 0mm 7mm 10mm, clip, width=0.6\linewidth]{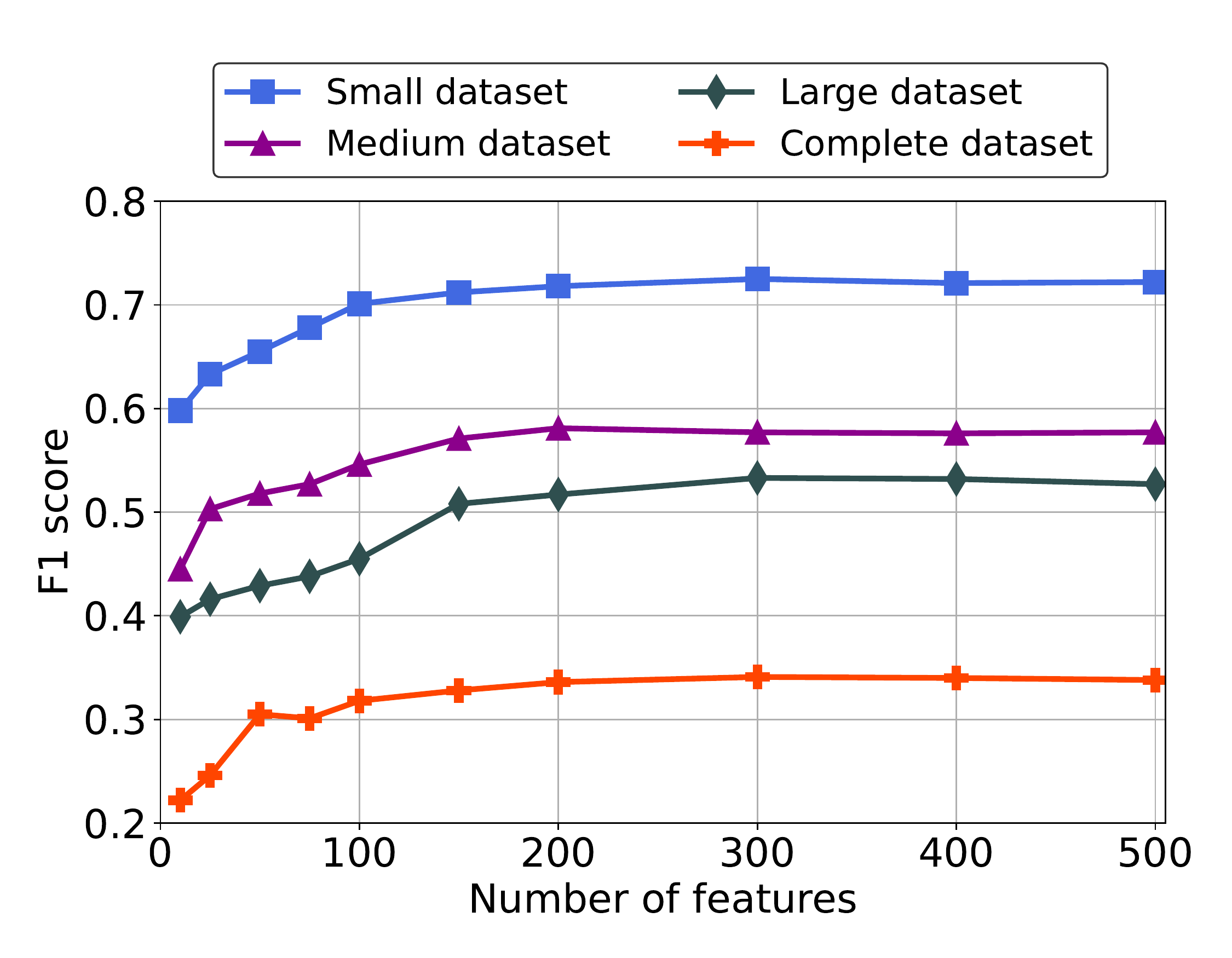}
			\label{fig:nof_multi}}
		\caption{Effect of $NoF$ value on F1-score.}
		\label{fig:nof}
	\end{figure}

	As elaborated in Section~\ref{subsection:feature_extraction}, 
	the tsfresh library provides nearly 1500 features from the extracted tail and delta TS. To retain only critical features, we first fix the maximum $NoF$ to be used in a given classification task. To find such an upper limit on $NoF$, we empirically observe the effect of $NoF$ values on the classification quality. As depicted in \figurename~\ref{fig:nof}, the F1-scores of binary and multi-class classifications stabilize near $NoF=100$ and $NoF=200$, respectively. Thus, we use $NoF=100$ for tasks with two classes and $NoF=200$ for tasks with multiple classes. A higher value of $NoF$ in multi-class classification indicates that it requires more features to distinguish among an increased number of EV classes.
\end{document}